\documentstyle[12pt]{article}
\begin{document} 
\begin{flushright} {OITS 666}\\
January 1999
\end{flushright}
\vspace*{1cm}

\begin{center}  {\large {\bf Erraticity Analysis of Soft Production by
ECOMB}}
\vskip .75cm
 {\bf  Zhen Cao$^{(1)}$ and Rudolph C. Hwa$^{(2)}$}
\vskip .3cm
 $^{(1)}$High Energy Astrophysics Institute, Department of
Physics,\\ University of Utah, Salt Lake City, UT 84112\\
  $^{(2)}$Institute of Theoretical Science and Department of
Physics\\ University of Oregon, Eugene, OR 97403-5203
\end{center}

\begin{abstract}
Event-to-event fluctuations of the spatial patterns of the final states of
high-enery collisions, referred to as erraticity, are studied for the data
generated by a soft-interaction model called ECOMB.  The moments
$C_{p,q}$ do not show simple power-law dependences on the bin size.  New
measures of erraticity are proposed that generalizes the bin-size
dependence.  The method should be applied not only to the soft production
data of NA22 and NA27 to check the dynamical content of ECOMB, but also 
to other collision processes, such as $e^+e^-$ annihilation and heavy-ion
collisions.

\end{abstract}

\section{Introduction}

Progress in the study of multiparticle production has recently been
made in two distinct directions among many others.  One is in finding
measures of event-to-event fluctuations \cite{1} that can probe the
production dynamics more deeply than the conventional observables,
such as multiplicity distribution and factorial moments \cite{2}.  Such
measures have been referred to as erraticity \cite{3}, which
quantifies the erratic nature of the event structure.  The other
direction is in the construction of a Monte Carlo generator, called
ECOMB \cite{4}, that simulates soft interaction in hadronic collisions,
capable of reproducing the intermittency data \cite{5}.  ECOMB stands
for eikonal color mutation branching, which are the key words of a
model that is based on the parton model rather than the string model
for low $p_T$ processes.  In this paper we combine the two, using
ECOMB to generate events from which we calculate the erraticity
measures.  The result should be of considerable interest, since, on the
one hand, the erraticity analysis of the NA22 data \cite{5} is currently
being carried out, and, on the other, it can motivate the investigation
and comparison of erraticities in various different collision processes,
ranging from $e^+e^-$ annihilation to heavy-ion collisions.

The study of erraticity originated in an attempt to understand
possible chaotic behaviors in quark and gluon jets \cite{1}, since QCD is
intrinsically nonlinear.  In the search for a measure of chaos it was
realized that the fluctuation of the hadronic final states of a parton jet
is the only observable feature of the QCD process that can replace the
unpredictable trajectories in classical nonlinear dynamics.  A
multiparticle final state in momentum space is a spatial pattern.  Once
a measure is found to quantify the fluctuation of spatial patterns, the
usefulness of the method goes far beyond the original purpose of
characterizing chaoticity in perturbative QCD processes.  Many
problems involve spatial patterns; they can range from phase
transition in condensed matter to galactic clustering in astrophysics. 
Even continuous time series can be transformed by discrete mapping
to spatial patterns \cite{6}.  Thus the erraticity analysis, which is the
study of the fluctuation of spatial patterns, is more general than the
determination of chaotic behavior.  Indeed, we have applied it to the
study of phase transition in magnetic systems by use of the Ising
model \cite{6}, as well as to the characterization of heartbeat
irregularities in ECG time series \cite{7}.

Multiparticle production at low $p_T$ has always eluded
first-principle calculation because of its nonperturbative nature. 
Various models that simulate the process can generate the average
quantities, but fail in getting correctly the fluctuations from the
averages \cite{2}.  In particular, few models can fit the intermittency
data \cite{5}.  To our knowledge ECOMB is the only one that can
reproduce those data \cite{4}, (apart from its predecessor ECCO
\cite{8}).  Since that model is tuned to fit the data by the adjustment
of several parameter, it is necessary to test its predictions on some
new features of the production process.  Erraticity is such a feature. 
The fluctuation of final-state patterns presents a severe test of any
model.

ECOMB includes many sources of fluctuations in hadronic collisions.  In
the framework of the eikonal formalism it allows for fluctuations in
impact parameter $b$.  For any $b$ there is the fluctuation of the
number $\mu$ of cut Pomerons.  For any $\mu$ there is the
fluctuation of the number $\nu$ of partons. For any $\nu$ the color
distribution along the rapidity axis can still fluctuate initially.  During
the evolution process the local subprocesses of color mutation, spatial
contraction and expansion, branching into neutral subclusters, and
hadronization into particles or resonances can all fluctuate.  Taken
together the model can generate such widely fluctuating events that
fitting some average quantity such as $\left< n\right>$ or $dn/dy$
does not explore the full extent of its characteristics.  The dependence
of normalized factorial moments $F_q$ on the bin size $\delta$
usually called intermittency, probes deeper, but it is nevertheless a
measure that is averaged over all events.  Erraticity is a true measure
of event-to-event fluctuation.

\section{Erraticity}

There are various ways to characterize a spatial pattern.  We shall use
the horizontal factorial moments.  Given the rapidity distribution of a
particular event, we first convert it to a distribution in the cumulative
variable $X$ \cite{9,1}, in terms of which the average rapidity
distribution
$dn/dX$ is uniform in $X$.  We then calculate from that
distribution for that event the normalized
$F_q$ 
\begin{eqnarray} 
F_q = \left<n (n-1)\cdots (n-q+1) \right> / \left<n
\right>^q \quad ,
\label{1}
\end{eqnarray} where $\left<\cdots \right>$ signifies (horizontal)
average over all bins, and $n$ is the multiplicity in a bin.  We
emphasize that (1) does not involve any average over events.  $F_q$
does not fully describe the structure of an event, since at any fixed
$q$ it is insensitive to the rearrangement of the bins.  However, it does
capture some aspect of the fluctuations from bin to bin, and is
adequate for our purpose.

Since $F_q$ fluctuates from event to event, one obtains a (vertical)
distribution
$P(F_q)$ after many events.  Let the vertical average of $F_q$
determined from
$P(F_q)$ be denoted by $\left< F_q\right>_v$ .  Then in terms of the
normalized moments for separate events
\begin{eqnarray}
\Phi_q = F_q/\left< F_q\right>_v \quad ,
\label{2}
\end{eqnarray} 
we can define the vertical $p$th order moments of the normalized
$q$th order factorial (horizontal) moments
\begin{eqnarray} 
C_{p,q} = \left<\Phi^p_q\right>_v  \quad .
\label{3}
\end{eqnarray} 
Erraticity refers to the power law behavior of
$C_{p,q}$ \cite{1,3}
\begin{eqnarray}
 C_{p,q}  \propto M^{\psi_q (p)} \quad ,
\label{4}
\end{eqnarray} 
where $M$ is the number of bins, $1/\delta$, and the
length in $X$ space is 1.  $\psi_q(p)$ is referred to as the erraticity
exponent. If the spatial pattern never changes from event to event,
$P(F_q)$ would be a delta function at $\Phi_q = 1$, and $C_{p,q}$
would be 1 at all $M$, $p$, and
$q$, resulting in $\psi_q (p) = 0$.  The larger $\psi_q (p)$ is, the more
erratic is the fluctuation of the spatial patterns.

Since $\psi_q (p)$  is an increasing function of $p$ with increasing
slope, an efficient way to characterize erraticity with one number (for
every $q$) is simply to use the slope at $p = 1$, i.e.
\begin{eqnarray}
\mu_q=\left.{d\over dp}\psi_q(p)\right|_{p=1}.
\label{5}
\end{eqnarray} 
It is referred to as the entropy index \cite{1}. 
Experimentally, it is easier to determine first an entropy-like quantity
$\Sigma _q$ directly from $\Phi_q$:
\begin{eqnarray}
\Sigma _q = \left< \Phi_q \, {\rm ln} \Phi _q\right>_v \quad ,
\label{6}
\end{eqnarray} 
which follows from (\ref{3}) and 
\begin{eqnarray}
\Sigma _q = \left.
d   C_{p,q}/dp \right|_{p=1} \quad ,
\label{7}
\end{eqnarray}  
and then to determine $\mu_q$ from
$\Sigma _q$ using
\begin{eqnarray}
\mu_q= {\partial\, \Sigma_q \over \partial \, {\rm ln} M} \quad ,
\label{8}
\end{eqnarray}
provided that $C_{p,q}$ has the scaling behavior
(\ref{4}).  In \cite{1} it is found that $\mu_q$ is larger for quark jets
than for gluon jets, indicating that the branching process of the former
is more chaotic, or, in more words, the event-to-event fluctuation is
more erratic.

If the moments $C_{p,q}$ do not have the exact scaling behavior in $M$, as
in (\ref{4}), but have similar nonlinear dependences
on $M$, we can consider a
generalized form of scaling
\begin{eqnarray} 
C_{p,q}(M)\propto g(M)^{\tilde{\psi}(p,q)}\quad .
\label{9}
\end{eqnarray} 
If (\ref{9}) is approximately valid for a common $g(M)$ for all $p$
and $q$, it then follows from (\ref{7}) that
\begin{eqnarray}
\Sigma_q (M)\propto \tilde{\mu}_q\, {\rm ln} g(M) \quad ,
\label{10}
\end{eqnarray}
where
\begin{eqnarray}
\left.\tilde{\mu}_q = {d \over dp} \tilde{\psi} (p,q) \right|_{p=1}\quad.
\label{11}
\end{eqnarray} 
Despite the similarity between (\ref{5}) and (\ref{11}) ,
$\tilde{\mu}_q$ is distinctly different from $\mu_q$ and should not
be compared to one another unless $g(M)=M$.

If (\ref{10}) is indeed good for a range of $q$ values, then we expect a
linear dependence of $\Sigma_q$ on $\Sigma_2$ as $M$ is varied. 
Let the slope of such a dependence be denoted by $\omega_q$, i.e.,
\begin{eqnarray}
\omega_q = {\partial  \Sigma_q \over \partial  \Sigma_2 } \quad.
\label{12}
\end{eqnarray}
Then we have
\begin{eqnarray}
\tilde{\mu}_q = \tilde{\mu}_2\ \omega_q \quad.
\label{13}
\end{eqnarray} 
A variation of this scheme that makes use of an extra control
parameter $r$ in the problem is considered in \cite{6}.  It is found
there that the entropy indices determined that way are as effective as
Lyapunov exponents in characterizing classical nonlinear dynamical
systems.

\section{Scaling Behaviors}

The erraticity analysis described above involves only measurable
quantities, so it can be directly applied to the experimental data.  The
NA22 data at
$\sqrt{s} = 22$ GeV are ideally suited for this type of analysis, since
$F_q$ fluctuates widely from event to event \cite{5}.  The nuclear
collision data, such as those of NA49, can also be studied, but $p_T$
cuts should be made to reduce the hadron multiplicity to be analyzed,
thereby enhancing the erraticity to be quantified.

Here we apply the analysis to hadronic collisions generated by
ECOMB.  The parameters are tuned to fit $\left< n\right>$, $P_n$,
$dn/dy$ and $\left<F_q
\right>_v$ of the NA22 data \cite{5}.  Without any further adjustment
of the parameters in the model we calculate $C_{p,q}(M)$, which are
therefore our predictions for hadronic collisions at 22 GeV.  The results
from simulating $3\times 10^4$ Monte Carlo events are shown on the left
side of Fig. 1.  The lines are drawn to guide the eye.

From the points shown, it is clear that the dependences of $C_{p,q}$
on $M$ in the log-log plots are not very linear, especially for the more
reliable cases of $q = 2$ and $3$, where the statistics are higher.  Thus
the power-law behavior in (\ref{4}) is not well satisfied.  Since the
general behaviors of $C_{p,q}$ are rather similar in shape, we can
regard $C_{2,2}$ as the reference that carries the typical dependence
on $M$, and examine $C_{p,q}$ vs $C_{2,2}$ when $M$ is varied as an
implicit variable.  The results are shown on the right side of Fig. 1. 
We have left out the highest points that correspond to the smallest
bin size, since they show saturation at $q > 2$.  We have also left out the
points corresponding to ln\,$M=0$, since the scaling behaviors
do not extend to the biggest bin size. The straightlines are linear fits of the
points shown and lend support to the scaling behavior
\begin{eqnarray}
C_{p,q} \propto C^{\chi(p,q)}_{2,2} \quad .
\label{14}
\end{eqnarray} 
The slopes of the fits are $\chi(p,q)$, which are shown in Fig. 2.  One may
regard $\chi(p,q)$ as a representation of the erraticity properties of the
particle production data, when there is no strict scaling law as in
(\ref{4}).

The behavior of $\chi(p,q)$ exhibited in Fig. 2 can be described
analytically, if we fit the points by a quadratic formula for each $q$. 
The result is shown by the lines in Fig. 2.  Evidently, the fits are
excellent.  The properties of the smooth behaviors can be further
summarized by their derivatives at $p = 1$:
\begin{eqnarray}
\left. \chi^{\prime}_q \equiv {d \over dp} \chi(p,q) \right|_{p=1}\quad.
\label{15}
\end{eqnarray}
The values of $\chi^{\prime}_q$ are 0.834, 2.818, 5.243 and 7.847 for
$q=2,\cdots, 5,$ and are shown in Fig. 3.  We suggest that these values of
$\chi^{\prime}_q$ be used to compare with the experimental data.

Although $C_{p,q}(M)$ do not satisfy (\ref{4}), we can consider the
more general form (\ref{9}).  If the same function $g(M)$ is good
enough in (\ref{9}) for all $p$ and $q$, then it follows from (\ref{14})
that
\begin{eqnarray}
 \chi(p,q) = \tilde{\psi}(p,q) / \tilde{\psi}(2,2) \quad.
\label{16}
\end{eqnarray}
Using (\ref{11}) we then have 
\begin{eqnarray}
\tilde{\mu}_q = \tilde{\psi}(2,2)\chi^{\prime}_q  \quad.
\label{17}
\end{eqnarray}
It should be noted that, whereas $\chi^{\prime}_q$ follows only from the
scaling property of (\ref{14}), the determination of $\tilde{\psi}(2,2)$,
and therefore $\tilde{\mu}_q$, requires the knowledge of $g(M)$ in
(\ref{9}).

To determine $g(M)$, we write it in the form
\begin{eqnarray}
{\rm ln}\, g(M) = ({\rm ln}\, M)^a \quad .
\label{18}
\end{eqnarray}
By varying $a$, we can find a good linear behavior of ${\rm ln}\,C_{2,2}$
vs ${\rm ln}\, g(M)$, as shown by the dashed line in Fig. 4 for $a = 1.8$. 
The corresponding value of $\tilde{\psi}(2,2)$ determined by the slope of
the straightline fit is 0.119.  Using that in  (\ref{17}) yields a set of values of
$\tilde{\mu}_q$, which are shown in Fig. 5 by the open-circle points.  In
particular, we have
\begin{eqnarray}
\tilde{\mu}_2 = 0.099 \quad ,
\label{19}
\end{eqnarray}
a quantity that has a separate significance below.

We remark that in checking the validity of (\ref{9}) for values of $p$ 
and $q$ other than 2, one can improve the linearity of the points for each
$p$ and $q$ by slight adjustments of the value of
$a$.  If there is a range of possible
$g(M)$ that depends on $p$ and $q$ to yield the best fits,
however small the variations in $a$ may be, the scheme
 defeats the point of defining a
universal $\tilde{\psi}(p,q)$.  We thus propose that the emphasis of
the erraticity analysis should be placed on (\ref{14}),  which is independent
of $g(M)$, and that (\ref{9}) is examined only for $p = 2$, $q = 2$ so that
(\ref{17}) can be evaluated.

Since $\tilde{\mu}_q$ is distinct from $\mu_q$, we cannot compare
our result on $\tilde{\mu}_q$ with the theoretical values of $\mu_q$ found
for quark and gluon jets \cite{1}, nor with the experimental values of
$\mu_q$ determined from
$pp$ collisions at 400 GeV/c (NA27) \cite{10}.

The values of  $\tilde{\mu}_q$ can also be determined independently
by use of $\Sigma_q(M)$.  From the definition in (\ref{6}) we have
calculated $\Sigma_q$ as functions of ln\,$M$, as shown in Fig.\ 6(a).  Not
surprisingly, the dependences are not linear.  However, when $\Sigma_q$ is
plotted against $\Sigma_2$ in Fig.\ 6(b), they all fall into straightlines,
except for the point corresponding to the smallest bin for $q=5$ (which we
have left out for the fit).  The slopes, which give 
$\omega_q$ defined in (\ref{12}), are 
1.0, 3.244, 6.0, and 9.101 for $q=2,\cdots, 5$.  They are
shown in Fig.\ 7.  If we examine (\ref{10}) for $q = 2$ only, and 
plot $\Sigma_2$ vs ln\,$g(M)$ with $a = 1.8$, as in Fig.\ 4, we
obtain a linear behavior with a slope 
\begin{eqnarray}
\tilde{\mu}_2 = 0.095 \quad .
\label{20}
\end{eqnarray}
This value is to be compared with that in (\ref{19}) with only 4\%
discrepency.  Of the two methods of determining
$\tilde{\mu}_2$, this latter approach is more reliable, since the
derivative in $p$ at $p = 1$ is done analytically in the definition of
$\Sigma_q$ in (\ref{7}), whereas in the former approach the
differentiation is done in (\ref{15}) using the fitted curve in Fig. 2. 
Substituting (\ref{20}) into (\ref{13}), we can determine the values of
$\tilde{\mu}_q$ for $q>2$ from the values of $\omega_q$ in Fig. 7.  The
result is shown by the solid points  in Fig. 5.  Clearly, the two methods yield
essentially the same result.

Another way to check the degree of consistency of the two methods,
independent of  the details on $g(M)$, is to examine the ratio
$r_q=\chi'_q/\omega_q$. The quantities in that ratio are derived from the
straightline fits of ln\,$C_{p,q}$ vs ln\,$C_{2,2}$  and $\Sigma_q$ vs
$\Sigma_2$  (as in Fig.\ 1 and Fig.\ 6) without resorting to such equation as
(18).  According to (13) and (17), the ratio $r_q$ should be a constant,
independent of $q$. From the values of
$\chi'_q$ and $\omega_q$ given above in connection with Figs.\ 3 and 7,
we find that 
$r_q = 0.834, 0.867, 0.874$, and 0.862 for $q=2,3,4,5$. The
average is 0.86, so the standard deviation is at the 1-2\% level.  Evidently,
the two methods are quite consistent, whatever $g(M)$ may be. From (13)
and (17), one would expect $r_q$ to be $\tilde\mu_2/\tilde\psi(2,2)$,
which according to the numbers given in Fig.\ 4, is 0.798. The discrepency
from 0.86 is nearly 7\%. Thus the disagreement of the
values of $\tilde\mu_q$ in Fig.\ 5, though not large, has the same root as
the disagreement between (19) and (20), namely, the necessity to use a
specific form of $g(M)$. Nevertheless, at the level of inaccuracy of 4\%,
which is comparable to the typical uncertainty in the experimental data, the
value of 
$\tilde\mu_2$ given by either (19) or (20) clearly provides an effective
measure of erraticity in soft production. 

\section{Conclusion}

In conclusion, we recapitulate the two essential points of this paper. 
One is the prediction of ECOMB on the nature of fluctuations of the
factorial moments $F_q$ from event to event.  The other is the proposed
method of summarizing the scaling behaviors of $C_{p,q}$ that do not have
strict power-law dependences on the bin size.  The two aspects of this paper
converge on the new erraticity measures $\chi(p,q)$, $\chi^{\prime}_q$,
$\omega_q$ and $\tilde{\mu}_q$.

It is hoped that the data from both NA22 and NA27 can be
analyzed in terms of these measures so that the dynamics of soft
interaction contained in ECOMB can be checked by the experiments.

The proposed measures of erraticity are, of course, more general than
the application made here to soft production.  Event-to-event
fluctuation has recently become an important theme in collisions of all
varieties:  $e^+e^-$ annihilation, leptoproduction, hadronic collisions at
very high energies where hard subprocesses are important, and
heavy-ion collisions.  What was lacking previously is an efficient
measure of such fluctuations.  The erraticity measures proposed in
\cite{1,3}, now generalized to $\chi(p,q)$, $\chi^{\prime}_q$, $\omega_q$
and $\tilde{\mu}_q$ are well suited for that purpose.  They may be
redundant, if strict scaling in $M$ is good enough to give the
erraticity indices $\psi(p,q)$.  The method of treating less-strict scaling
properties proposed here may well be more generally
applicable to the wide range of collision processes amenable to erraticity
study. 

\vspace*{1cm}
\begin{center}
\subsubsection*{Acknowledgments}
\end{center}

This work was supported in part by U.S. Department of Energy under
Grant No. DE-FG03-96ER40972 and by the National Science Foundation
under contract No. PHY-93-21949.

\newpage
\begin{center}
\section*{Figure Captions}
\end{center}
\begin{description}

\item[Fig.\ 1]\quad Log-log plots of $C_{p,q}$ versus $M$ on the left side
and versus $C_{2,2}$ on the right side.  The lines on the left side are to
guide the eye, while the ones on the right side are linear fits.

\item[Fig.\ 2]\quad The slopes of the linear fits on the right side of Fig.\ 1
are plotted against $p$ for various values of $q$. The lines are fits by
quadratic formula.

\item[Fig.\ 3]\quad The derivatives of $\chi(p,q)$ in Fig.\ 2 at $p=1$.

\item[Fig.\ 4]\quad The open circles are for $C_{2,2}$ and the solid points
are for $\Sigma_2$. The lines are linear fits, whose slopes are
$\tilde\psi(2,2)$ and $\tilde\mu_2$, respectively.

\item[Fig.\ 5]\quad $\tilde\mu_q$ determined in two different ways: Eq.\
(17) for the open circles and Eq.\ (13) for the solid points.

\item[Fig.\ 6]\quad (a) $\Sigma_q$ vs ln $M$ for various $q$; (b)
$\Sigma_q$ vs $\Sigma_2$ with the lines being linear fits.

\item[Fig.\ 7]\quad The slopes of the straightlines in Fig.\ 6(b),
$\omega_q$, plotted against $q$.

\end{description}

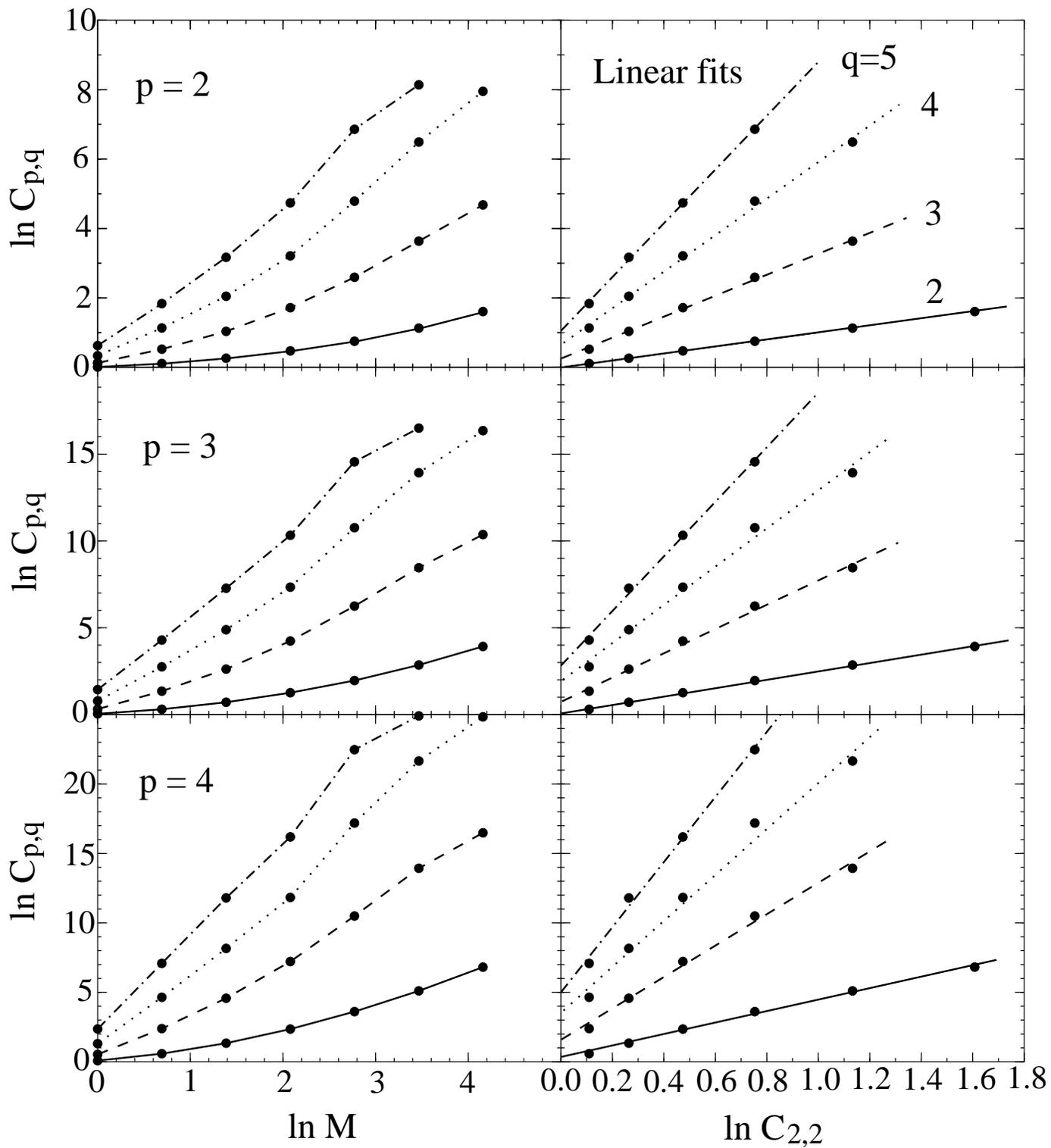
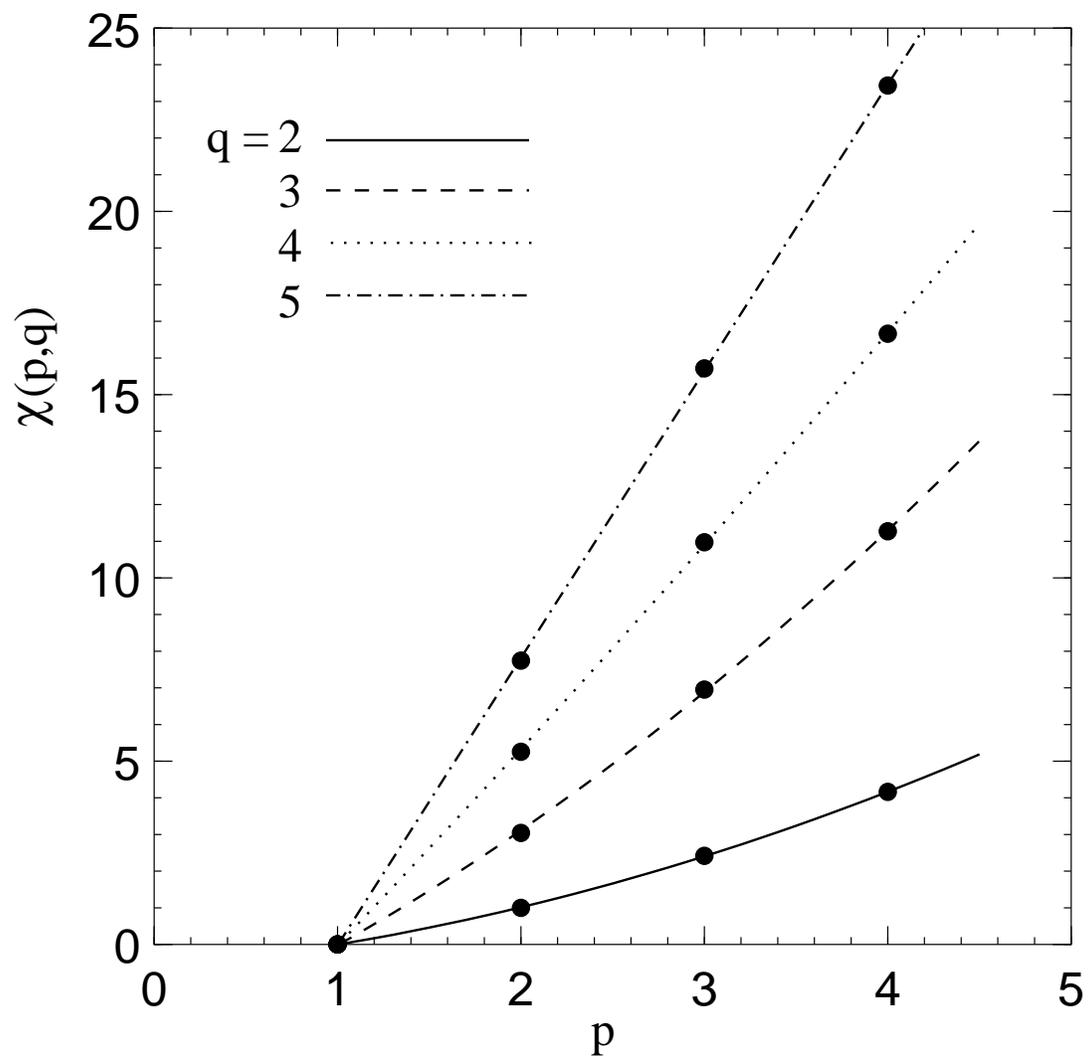
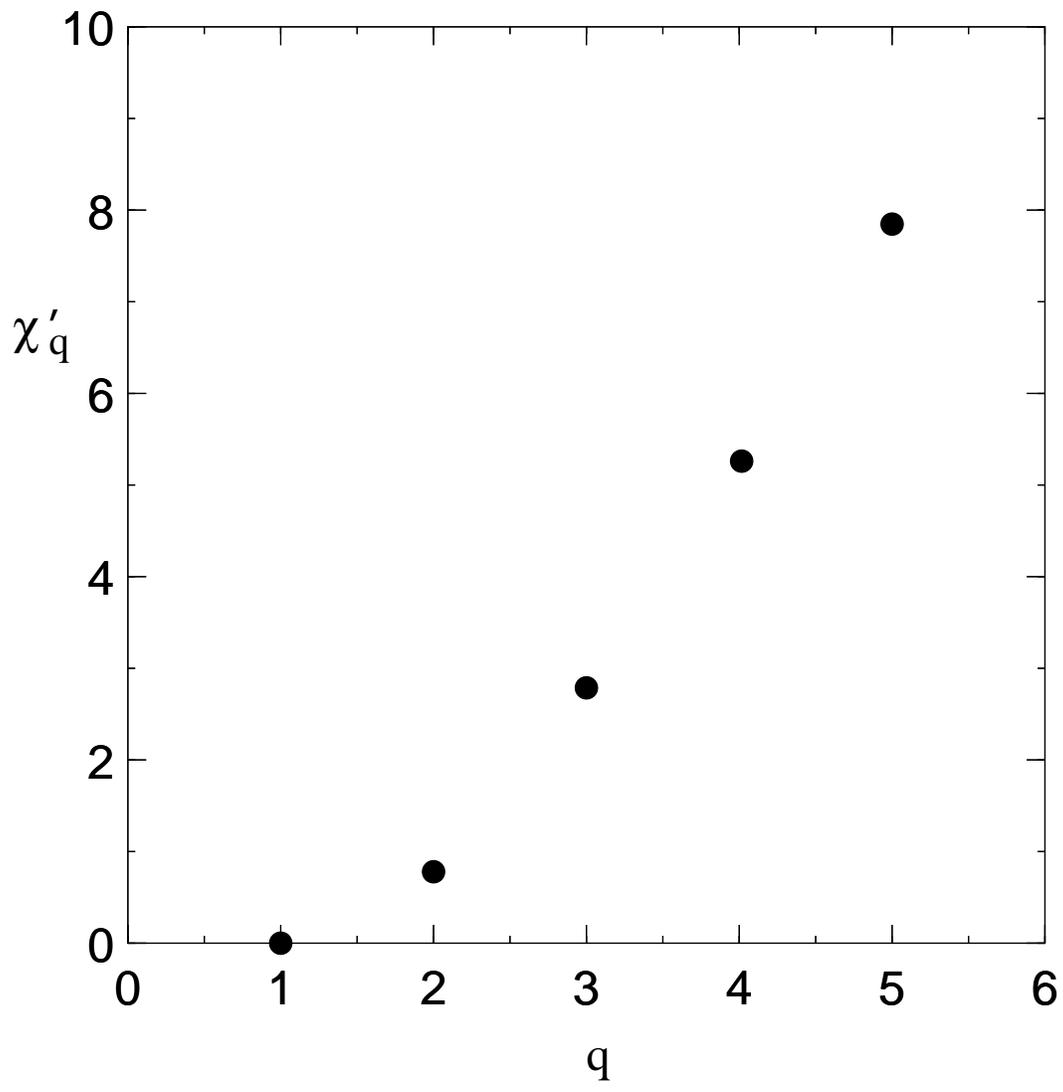
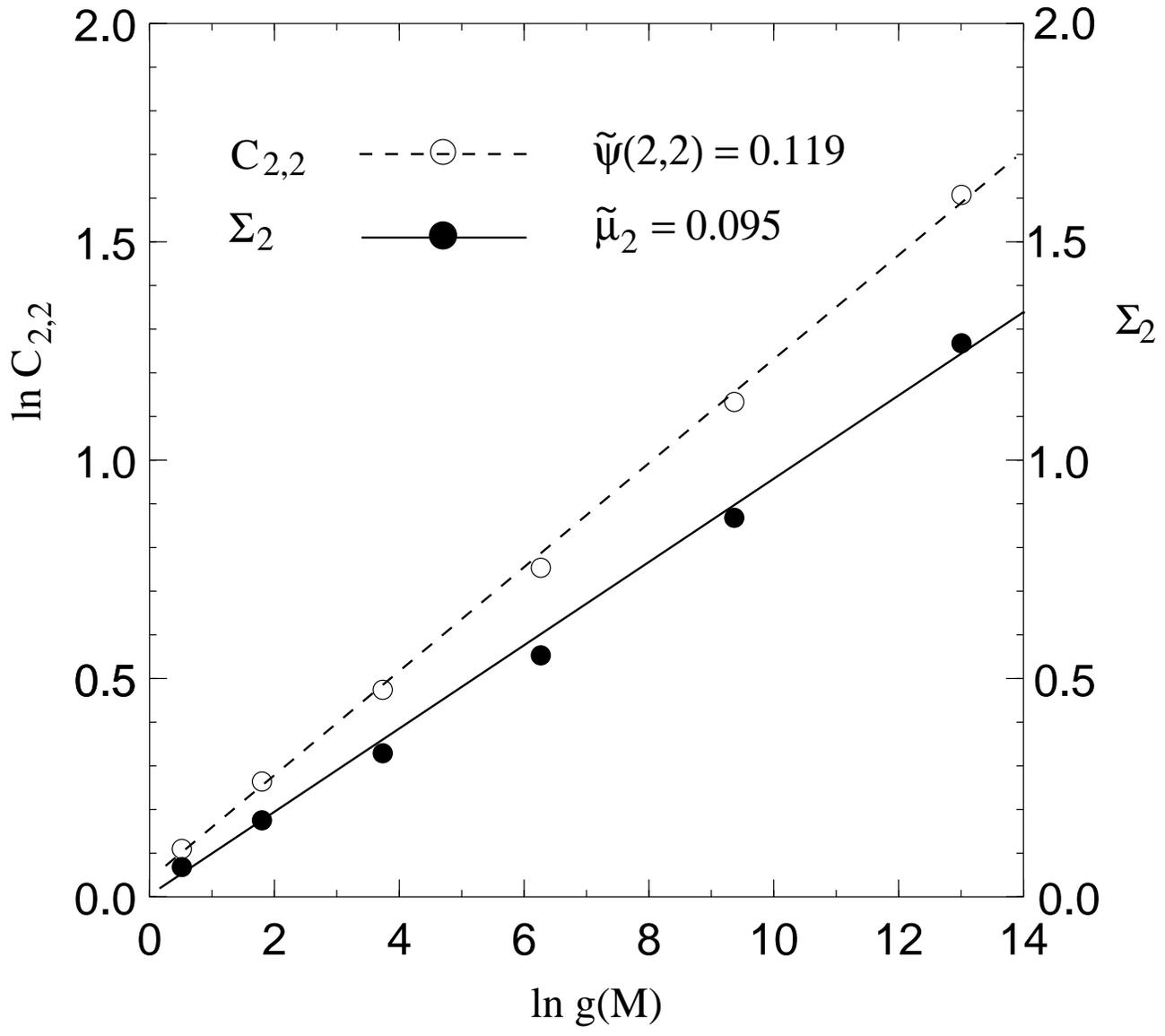
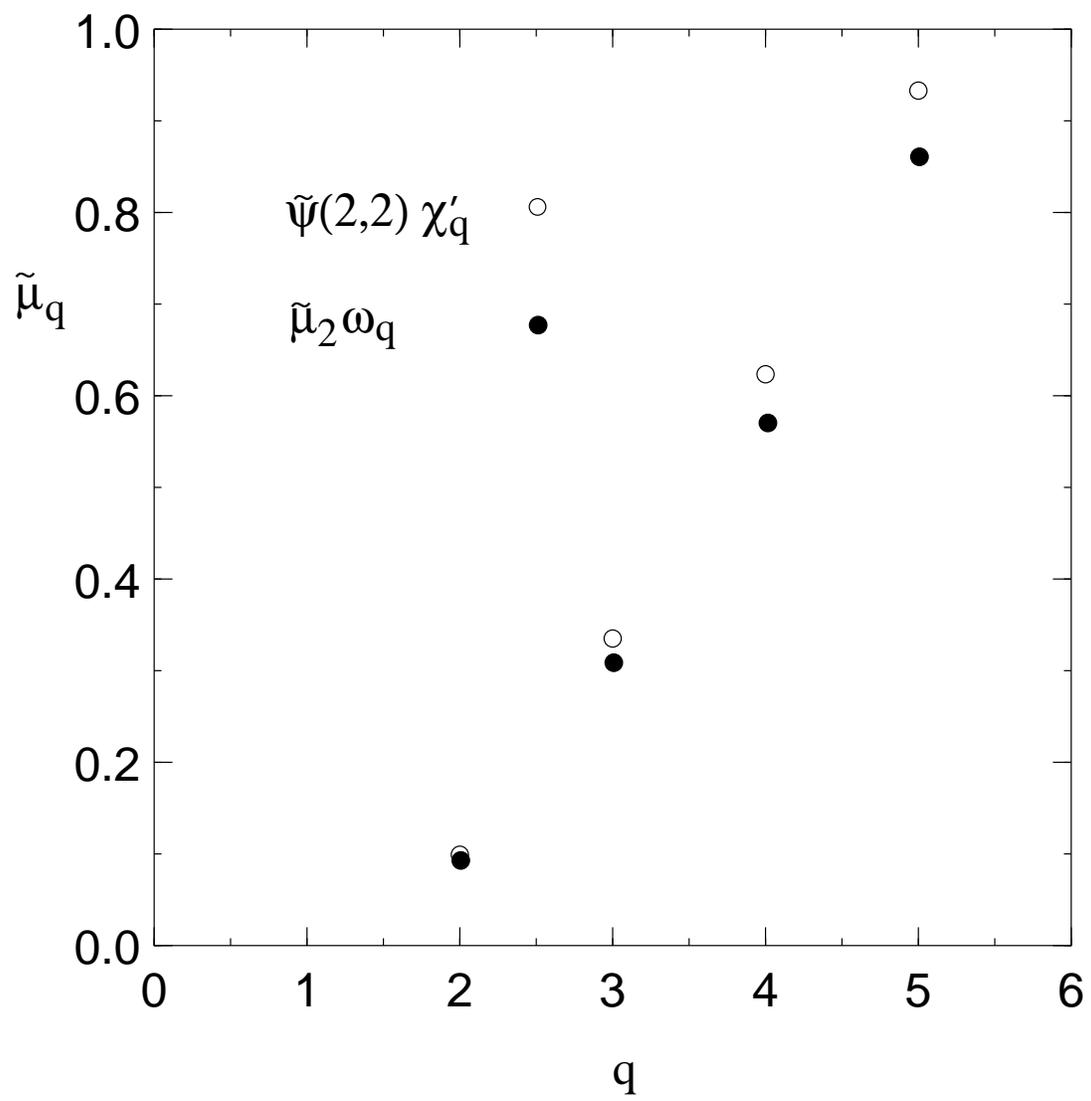
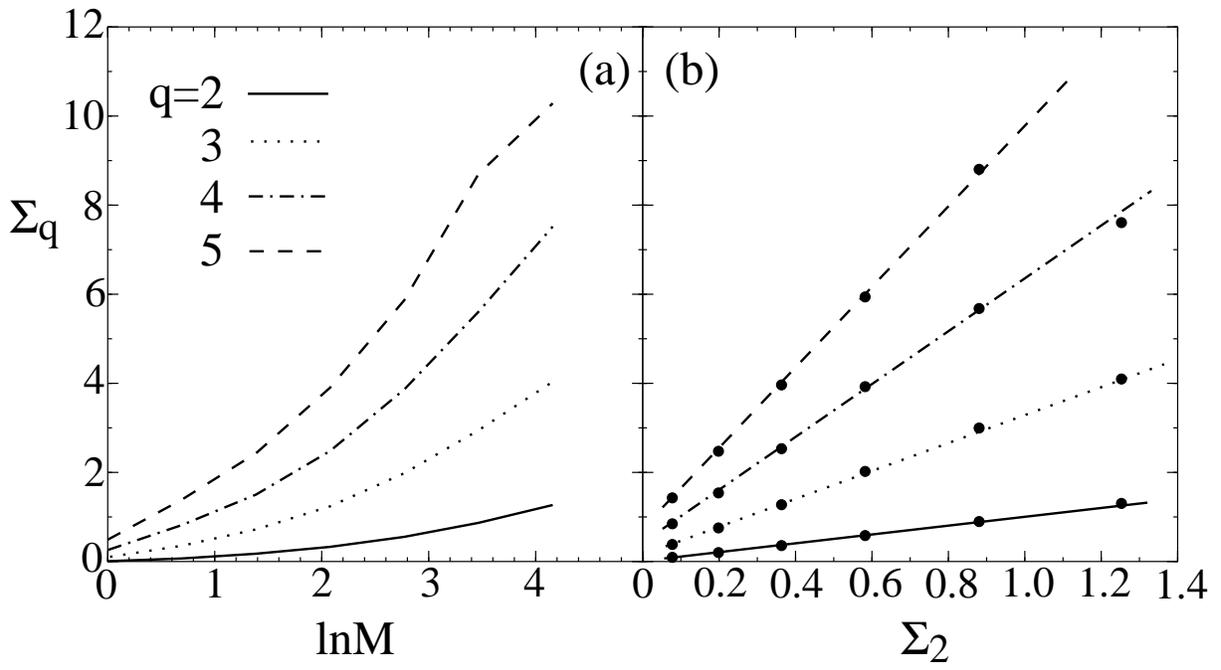
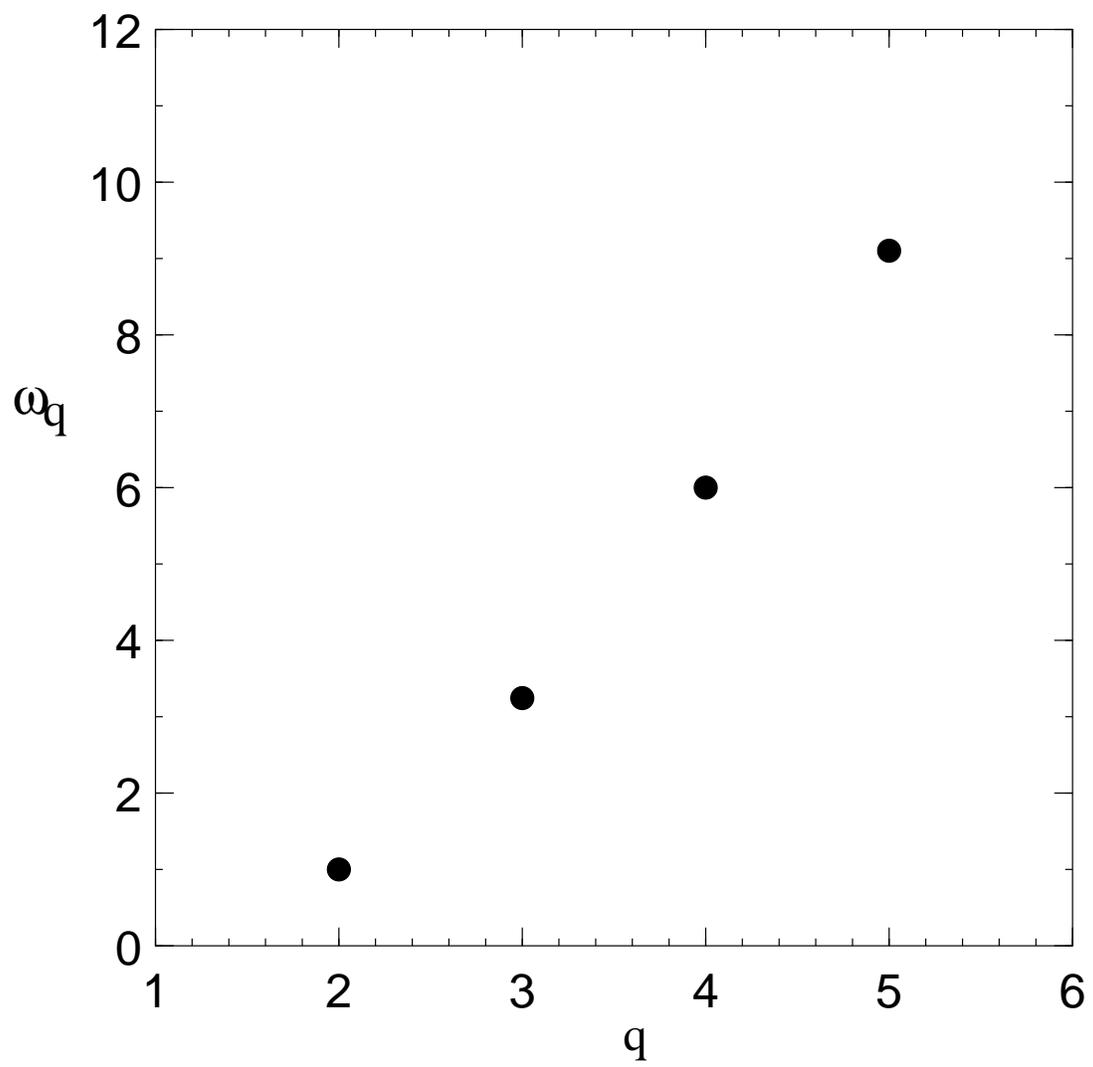
\end{document}